# Variability-Aware Design for Energy Efficient Computational Artificial Intelligence Platform

*Rhonda P. Zhang, Jason Yuan Du*

## Introduction

Portable computing devices, which include tablets, smart phones and various types of wearable sensors, experienced a rapid development in recent years. One of the most critical limitations for these devices is the power consumption as they use batteries as the power supply. However, the bottleneck of the power saving schemes in both hardware design and software algorithm is the huge variability in power consumption. The variability is caused by a myriad of factors, including the manufacturing process, the ambient environment (temperature, humidity), the aging effects and etc. As the technology node scaled down to 28nm and even lower, the variability becomes more severe. As a result, a platform for variability characterization seems to be very necessary and helpful.

In this paper, a platform to characterize the performance of a customized ARM Cortex M3 Processor was developed. Cortex M3 Processor is a widely used 32-bit processor in embedded systems for its low power consumption, low costs and high performance. This chip was jointly designed by University of Michigan and UCLA with ARM IP. On this platform, researchers have the full control of supply voltage and clock frequencies; while they can monitor the power consumption in active and sleep mode, leakage current and clock frequency. Simultaneously, with the ultra-scaling CMOS technology, higher and higher data bandwidth, big data analysis and energy-efficient system-on-chips (SoCs) become ubiquitous, in the applications including, smart wireless systems [11-20], navigation systems [21], smart human-computer interface systems [22-23], and cognitive communication systems [24-37].

In this paper, we will first give a brief description of the previous work done by University of Michigan; then, we will talk about the features of Cortex M3 processor; after that, we will discuss our motivation, the design of each part, including power regulator, frequency and current sensing and micro-controller; finally, we will summarize everything into a system level design and draw a conclusion.

## Previous Works

Michigan University designed a proto-type board which can be controlled via a PC. They created a visual system in National Instrument's Labview. It has the following functions:

1) Compile C code into hex file compatible with the M3 processor.

2) Control the supply voltage and provide the required power up, power down sequence.

3) Configure the PLL and internal clock frequency.

4) Read and write to SRAM via JTAG.

Their board is connected to a PC via PCI for analog I/O and PCI-E for digital I/O. It has two power supply modes. The first one is to provide +5V/-5V to the analog buffer on board, and the Labview program provides the reference voltage via analog I/O for the buffer input. In this way, the power up sequence is achieved by controlling the timing of the reference voltage in Labview. The other option is to use the Labview to control the programmable power supply directly, and connect the power supply output to the board via SMA connector. Two modes can be switched swiftly by the on board switches. Our replica of their platform chooses the first mode for its lost cost.

## ARM Cortex M3 Processor

The ARM Cortex M3 Processor is a very popular 32-bit RISC (Reduced Instruction Set Computer). University of Michigan and UCLA jointly purchased the license and customized it. Figure 1 shows the architecture of the processor. Several blocks were removed (indicated by the red crosses).The core doesn't have the interrupt functionality due to the absence of the NVIC [2].

In this configuration, the core is connected to the SRAM via Bus Matrix. The SRAM is accessible from outside by JTAG via DAP (Debug Access Port). More specifically, the debug mode can be enabled by writing certain values to the Debug Halting Control and the Status Register (DHCSR) via JTAG [12].

The internal clock can be switched from an external reference clock HCLK and the output of on-chip PLL. The on-chip PLL needs a 20MHz external reference clock. The control of PLL and other on-chip peripherals is via a customized scan-chain. The scan-chain is a 226-bit serial interface having both RESET and CORE_RESET, the former is to reset the whole chip whereas the latter only reset the ARM core.

The differences between the chips customized by Michigan and UCLA are shown below:

1) Michigan chip does not have the on-chip frequency and leakage sensors.

2) The scan chain of UCLA chip is one bit shorter than the Michigan chip. The missing bit is CORE_RESET. Hence, the only way to reset the core of UCLA chip is to write to CORE_RESET register 0X4400 0004 via JTAG.

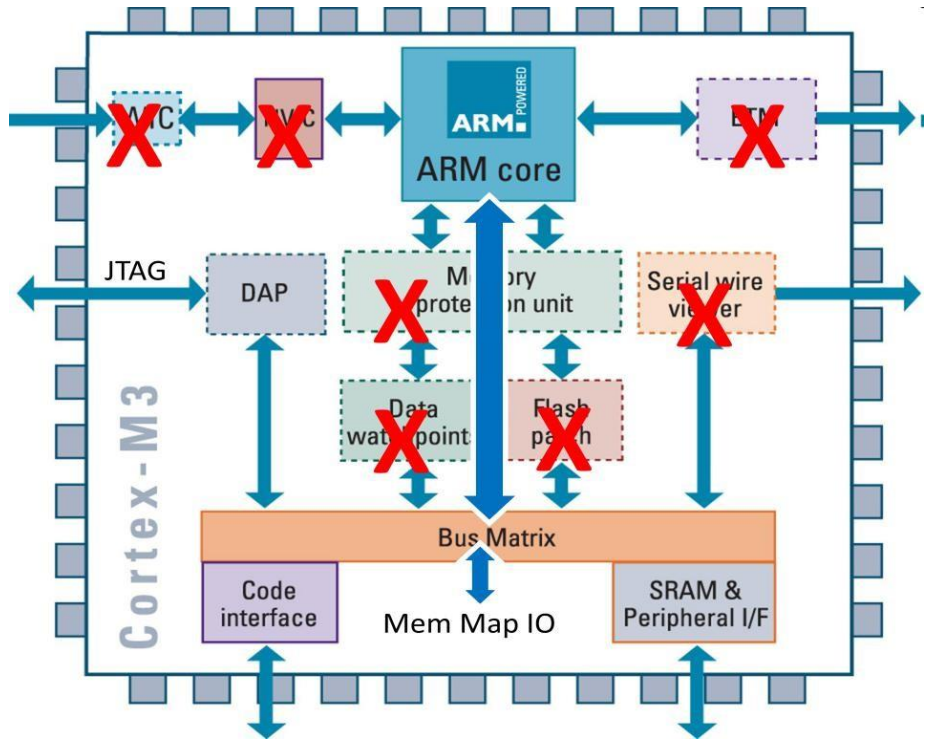

Figure 1: Architecture of the customized ARM Cortex-M3 Processor [2]

## Motivation

Monitoring the performance and power consumption of a processor is not a new idea. Several research projects and commercial products have adopted this approach to improve the energy efficiency. Here two of such examples will be described. One is the IBM POWER6 Processor, and the other is the Low Power Energy Aware Processing (LEAP) project at UCLA.

Figure 2 shows the conceptual topology of the POWER6-based energy awareness system. POWER6 is a microprocessor designed by IBM with Power Instruction Set Architecture (ISA) for server and mainframe applications. It has on-chip thermal sensors and critical path monitors (CPM). It also implements an external current and voltage sensing circuit for energy measurement. The on-chip actuator is able to control the state and activities of the processor and its DRAM subsystem by pipeline throttling. Pipeline throttling means to limit the rate at which instructions are dispatched for execution. Sensors and actuators form a self-contained closed-loop system which can achieve thermal and power capping. For example, the sensor is capable of measuring the temperature and the power consumption of the system and giving feedback to the control system. If the temperature or the power consumption exceeds the maximal values allowed, the actuators will apply pipeline throttling to lower the core activities, and thus lower the temperature and power consumption [6, 7].

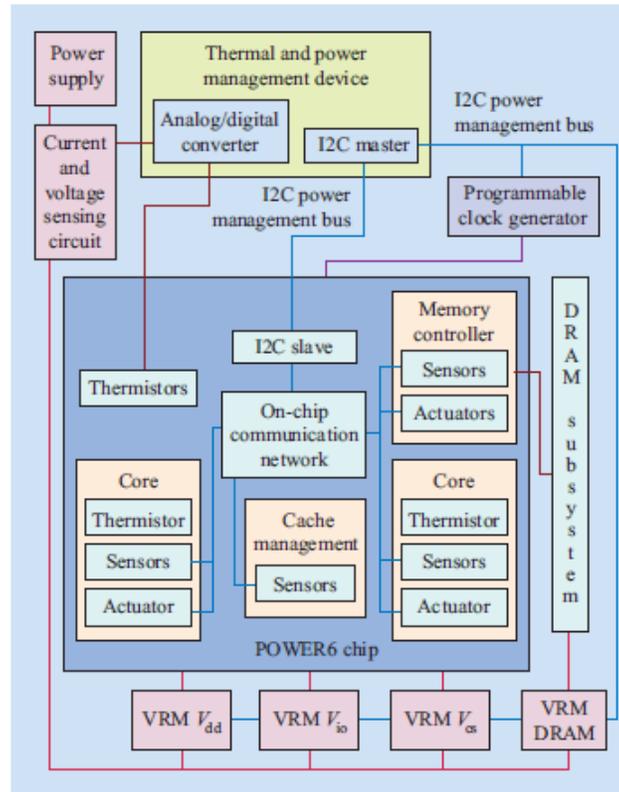

Figure 2: System-Level Block Diagram of IBM POWER6 [6]

LEAP is a combination of two subsystems, the Slauson Processor Module (SPM) in Figure 3 and the Energy Measurement and Accounting Preprocessor (EMAP) in Figure 4. It has a PXA255 processor in the SPM, which is an XScale architecture mobile application processor. SPM and EMAP are connected via a 180-pin inter-board header. The inter-board communication is via the I2C bus. The EMAP features a MSP430 micro-processor from Texas Instruments. The MSP430 is in charge of the following tasks: communication with SPM via I2C, power management scheduling, analog to digital conversion of sensor outputs and the control of radio CC2420 via SPI. THE EMAP can measure the power consumption of each power domain, and in return control the operation mode of the SPM [8, 9].

Leakage current shows a much greater statistical variations than active current. The leakage power increases dramatically and cannot be neglected anymore in virtue of the shrinking size of CMOS devices [10]. However, it is extremely difficult to monitor the leakage current due to its tiny amplitude, normally at the order of nanoamps. With these considerations in mind, this project aims to provide a platform which can measure not only the power consumption, but also the leakage current. Our chip features an on-chip leakage current sensor and a frequency sensor. The frequency sensor can provide an estimation of the maximum clock frequency.

Moreover, the customized Cortex-M3 core on the platform offers the possibility to characterize the variability of this popular embedded processor.

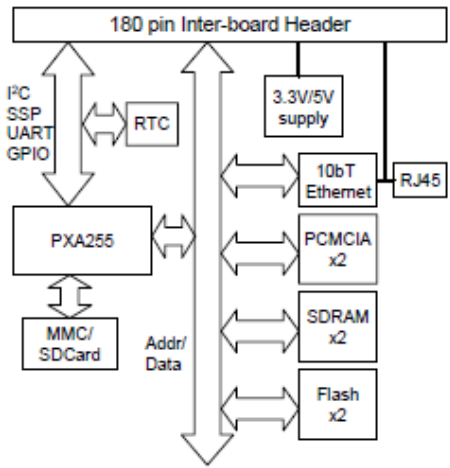

Figure 3: SMP Block Diagram [8]

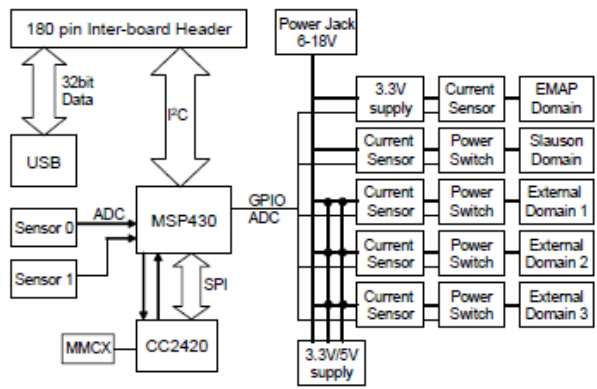

Figure 4: EMAP Block Diagram [8]

## Project Timeline

Since the project is a continuation of the project at University of Michigan, the project was started with replicating Michigan's test environment in order to reduce the design time. We soldered a Michigan's PCB (Printed Circuit Board) and used their Labview code to test their chip.

After understanding how their platform works, we designed our first PCB, the UCLA_C1B1, which is the first test board for the first UCLA fabricated chip. UCLA_C1B1 only brought out the frequency and leakage current sensors to pinheaders, so that we can test and characterize the performance of these sensors.

Then, we designed the UCLA_C1B2, which brought out all pins except those mounted sensors. We modified the Labview code to accommodate different features on the chips customized by UCLA and Michigan University, especially the scan chain part. We were able to use the Labview to drive the UCLA chip to run some simple test programs. The test procedure differs slightly as well due to the missing CORE_RESET bit in scan chain.

Finally, we designed the UCLA_C1B3, which is a closed-loop independent platform for variability characterization. On this platform, we have the control of supply voltages and clock frequency, while we can measure the leakage current, power consumption and the estimation of maximum clock frequency. The rest of this report will discuss in detail about the design for the UCLA_C1B3.

## Power Supply

Our customized Cortext-M3 processor supports multiple power domains, which are isolated and can be measured independently. Its operation requires eight levels of voltage supplies, i.e. DVDD (3.3V), DVDD2 (1.8V), AVDD (1.0V) and AVDD2 (0.5V) for the analog and digital peripherals, COREVDD for ARM core, SRAMVDD for on-chip SRAM, WRAPPERVDD for wrappers and SENSEVDD for on-chip sensors [1]. The voltage level of COREVDD and SRAMVDD need to be adjustable by software for voltage sweeping. We also connected a jumper in series with all other 6 supplies to leave the capability of current measurement. Besides, there is a special power up and power down sequence (figure 2 and figure 3) which should be full-filled.

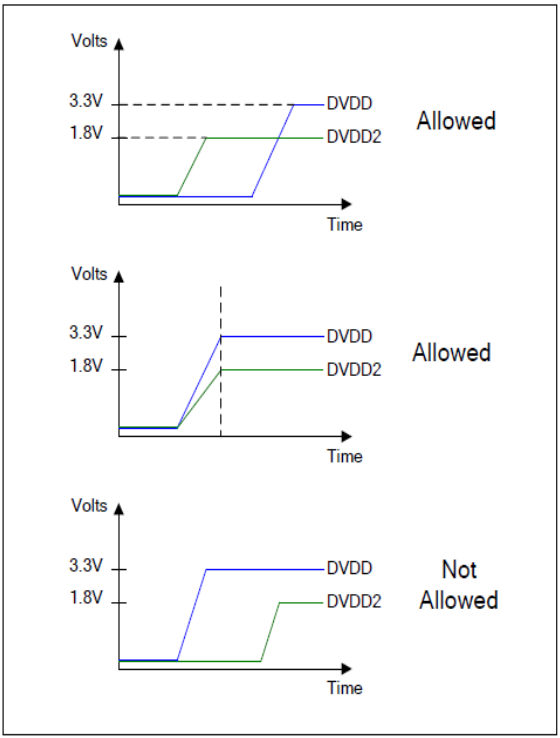 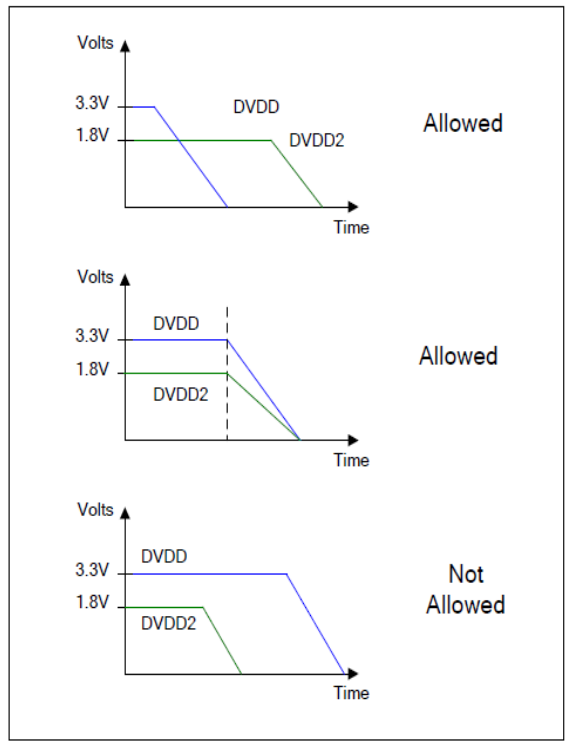

Figure 5: Power up sequence [1]                    Figure 6 : Power down sequence [1]

The possible approach to overcome these design constraints is to use dynamically adjustable regulators for COREVDD and SRAMVDD, and to use low dropout regulators (LDO) with Enable functions for the rest. However, the limitation is that most adjustable regulators (such as MAX8581) and LDOs with Enable (such as MAX8840 and LP38511) have only one channel. It complicates the design as at least two adjustable regulators, two LDOs with Enable, and one regulator with four channels are needed.

Inspired by Michigan's design, we decided to use the similar approach, namely providing reference voltage to the analog buffer. The reference voltages can be provided by a multi- channel Digital to Analog Convertor (DAC). The voltage sweep can be achieved by adjusting the DAC's outputs while the power sequence can be achieved by activating each channel in a sequential order. We used DAC7578 from Texas instruments in this project. It is an 8-channel 12-bit DAC with a I2C interface. The output voltage level and sequence can be programmed by a micro-controller via the I2C. Its output will be kept at GND until a valid value is written into its register [13]. The analog buffer used is EL5427 from Intersil, which has 12 channels each of which can provide up to 30mA current [14].

DAC7578 requires a 1kOhm to 10kOhm pull-up resistor on its two I2C lines, SDA and SCL. It also needs a 0.1uF and a 10uF bypass capacitor. These two capacitors should be placed closed to the DAC. In addition, a 470uF capacitor is needed to decouple low frequency noise.

# Current and Frequency Sensing

The on-chip frequency sensor can produce a 16-bit digital output from pin **O7** to **O0**. These pins are multiplexed between the higher 8 bits and the lower 8 bits by the output select pin **OS**. The sensor needs a counter clock signal and several control signals, including **READ**, **ENABLE**, **COUNT** and select signals **S5**-**S0**. **SELECT** is to select the ring oscillator (RO) under testing from 60 ROs on-chip. **ENABLE** is active high, and the counting starts after an impulse in **COUNT**. **READ** will send the counted values to the output. All of these pins should be connected to the digital I/O of a micro-controller.

The leakage current sensor has six pins. The outputs of **RVTP** and **RVTN** indicate the amplitude of the leakage current measured from regular threshold voltage ($V_{th}$) of PMOS and NMOS respectively; while **HVTP** and **HVTN** are the counterparts for high threshold voltage [3]. The gate and source of PMOS and NMOS of different types are connected together. **DEVICE_G** and **DEVICE_S** are the external connections to the gate and the source. The leakage current is very tiny, at the order of 10s of nanoampere. It needs to be amplified before sending to any Analog to Digital Converter (ADC). We use an integrator to amplify the signal. As this is the most difficult part to design on this platform, we leave a socket for daughter card in case the design on the mother board doesn't work properly. The daughter card can be plugged in perpendicularly to the motherboard by edge connector, which provides good signal shielding and area flexibility.

The power consumption of the ARM core and SRAM is approximated by measuring the current flown through the COREVDD and SRAMVDD. The current is calculated by measuring the voltage across a serial resistor. The supply voltage is as low as 0.6V, thus the voltage drop of the serial resistor should be smaller than 10mV. The current through COREVDD is as high as 15mA in the Active mode and as low as 1mA in the Sleep mode. To satisfy the voltage drop requirement, we used three 1ohm precision resistors in parallel for current sensing. It would give a 5mV voltage drop in the Active mode, and 0.3mV drop in the Sleep mode.

This voltage drop needs to be amplified at least by 200 before applied to any ADC. The large dynamic range makes the design challenging as most current sense amplifiers has comparable input offset voltage. We used INA333 from TI for the amplifier, which has a 25uV input offset. Its gain is adjustable by the equation $G= 1+100kOhm/R_G$, where $R_G$ is the resistance connected between its pin RG+ and RG- [15].

The design for SRAMVDD is much easier as its current is fairly stable around 1mA. We used a 7.5Ohm precision resistor and same current sense amplifier for the measurement. Both amplifiers were set to a gain of 200. To make the voltage measured and gain accurate, we used high precision resistors which have only 0.1% tolerance. A 0.1uF ceramic capacitor is placed very closely to the amplifier to decouple noises.

## Micro-controller

The micro-controller on our platform plays a similar role to that of the Labview in Michigan's design. It needs to do SRAM access via JTAG, PLL configuration via scan chain, control the DAC for power sequence, read and process the sensing values from various sensors. Mbed is our optimum choice due to its user-friendly environment and our past experience.

Two Mbeds are needed because of the large amount of pin connections. The Master Mbed is in charge of the JTAG, scan chain and power sequence; whereas the slave controls those sensors. The output from the leakage current sensor and the power sensors are connected to the Analog to Digital Converter (ADC) of the slave Mbed. Mbed has 10-bit ADC from pin 15 to pin 20. Two Mbeds are synchronized using the I2C bus. The Master Mbed also has the pin-headers to Ethernet for possible wireless applications in the future. There are only 26 General Purpose I/O (GPIO) on Mbed. Some of the pins on the Slave Mbed are multiplexed between the frequency sensing and leakage current sensing [16]. Table 1 in Appendix shows the detailed pin assignment of Mbeds.

## System-Level Design

Figure 7 shows the complete system-level design. Mbed 1 is the master, and Mbed 2 is the slave. Figure 8 highlights the floor planning of the UCLA_C1B3. We brought many pins to pin-headers for easy testing purposes.

Two Mbeds are placed at the boarder of the chip for easy USB plug-in. The DAC and analog buffer stay very closely to Mbed 1. It goes through the Current Sense Amplifier block before distributed to power supply network. Clock oscillators are placed on the left hand side of the chip as they are closed to the clock pins on the chip. This arrangement also avoids the long routes of high frequency clock signal, which could add unwanted noise to adjacent signals. The chip socket is placed in such an orientation so that the power and the clock signal can be accessed on the left-hand side, while leakage sensing outputs are on the right hand side. More importantly, the large open space in the center and the top offers much easier routing for the large amount of connections between the chip and Mbed2. The leakage current amplifier is placed on the right of the chip with the shortest possible routing because it is the most sensitive signals on board due to its tiny amplitude. The socket for daughter-card is placed right after the amplifier.

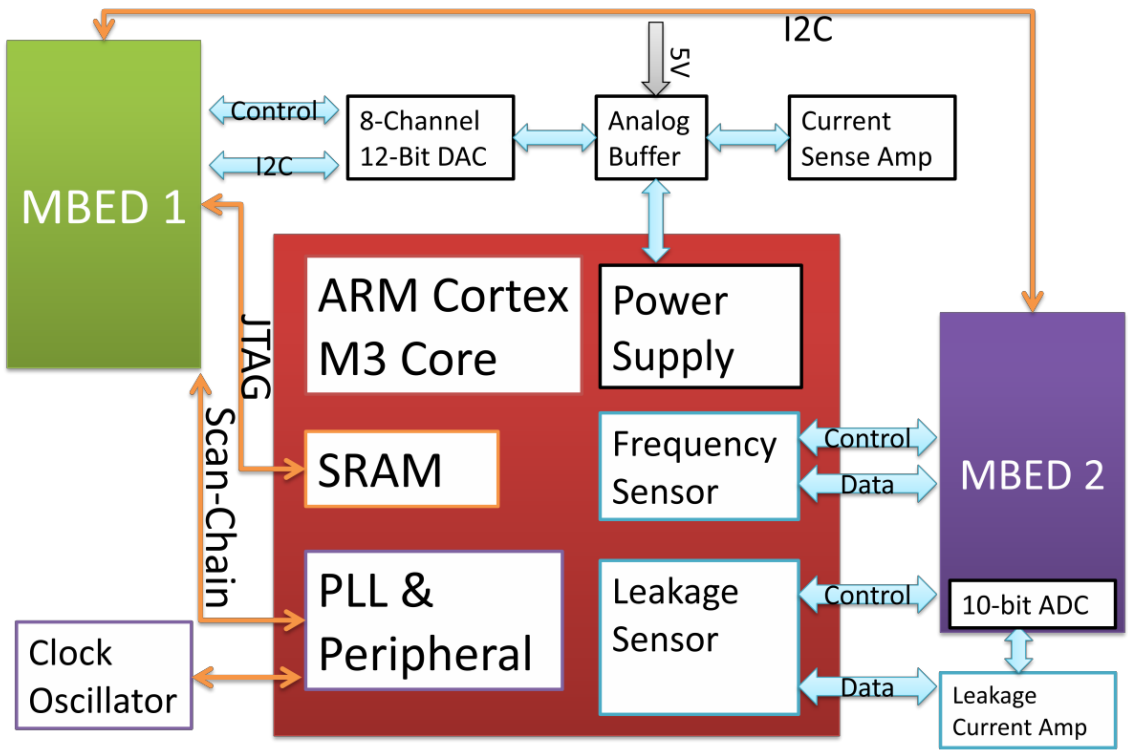

Figure 7: System-Level Block Diagram of UCLA_C1B3

All power supplies are decoupled with 470uF electrolytic capacitors for low frequency and 10uF ceramic capacitors for high frequency.

Most components use surface mounted package (SMD) to save space. We removed those unused pins in the ZIF socket to save further space for routing. Top layer (red) is routed horizontally while bottom is routed vertically in most cases. All power traces are 0.012 and 0.010 inch wide to minimize the voltage drop. All other signals are 0.006 inch wide.

The code sequence for two Mbeds should be:

1. Write to the DAC via I2C after system reset
2. Set the PLL via scan chain
3. Reset the chip via BASE_RESET pin
4. Enable the DAP via JTAG
5. Write to the DHCSR for debug mode via JTAG
6. Load the program into SRAM via JTAG
7. Set the MEM MAP bit by writing 1 to its register via JTAG
8. Do CORE_RESET by writing 1 to 0X4400 0004 via JTAG
9. Wait, and read the results from the SRAM via JTAG
10. Sync Slave with Master via I2C regularly

11. Measure the power consumption, leakage current and internal clock frequency on Slave in parallel with step 3 to step 9
12. Send the measured data from Slave to Master, Master record these data with the DAC and PLL setting

## Conclusion

To sum up, in this project we build a platform on PCB to characterize the variability of a customized ARM Cortex M3 Processor. On this platform, the clock frequency and supply voltage can be swept for variability characterization. It can measure the leakage current and power consumption in the Active and Sleep mode. It also offers the capability to estimate the maximum possible clock frequency which can be run on the chip by the on-chip frequency sensor.

The key feature of this platform is that it is an independent closed-loop system. During the whole testing process, the platform is self-contained and does not need any external intervention.

This board can also be placed into a temperature chamber to test the variability against temperature. However, to close the loop for the temperature sweeping, an on-board temperature sensor and some wireless control of the temperature chamber are required.

The possible future improvement for this project includes limiting the power consumption of external components and hardware resources used. Table 2 in the appendix shows the power budget of external components. The total power consumption is about 1.45W. Our estimation of the current drawn by the chip is 15mA, and thus its power is about 50mW. It is significantly lower than the external power consumption. To make the platform feasible for wider applications, we have to reduce the external power consumption.

Besides, the chip itself should incorporate more functionality, such as power supply regulation, leakage current amplification and etc.